\documentclass[aps,twocolumn,showpacs]{revtex4-1}
\usepackage{graphicx}        % Include figure files
\usepackage{dcolumn}        % Align table columns on decimal point
\usepackage{bm}                % bold math
\usepackage{xcolor}
\usepackage{ulem}

\usepackage{times,mathptmx}

\begin{document}

\title{Signature of the $s$-wave regime  high above ultralow temperatures}
%subtitle: influence of $s$-waves in resonant exchange scattering

\author{Robin C\^ot\'e and Ionel Simbotin}

\affiliation{Department of Physics, U-3046, University of Connecticut, Storrs, CT,
06269}

\date{\today}

\begin{abstract}

  Physical processes involving charge transfer, spin
  exchange, or excitation exchange often occur in conditions of
  resonant scattering.  We show that the $s$-wave contribution can be
  used to obtain a good approximation for the full cross section.
  This approximation is found to be valid for a wide range of
  scattering energies, including high above the Wigner regime, where
  many partial waves contribute.
    We derive an analytical expression for the exchange
  cross section and demonstrate its relationship to the Langevin cross
  section.  We give examples for resonant charge transfer as well as
  spin-flip and excitation exchange.  Our approximation can be used to gain information
  about the $s$-wave regime from data obtained at much higher
  temperatures, which would be advantageous for systems where the
  ultracold quantum regime is not easily reachable.
\end{abstract}

\pacs{}%34.70.+e, 31.15.ap, 31.50.-x}

\maketitle

In recent years, rapid progress has been made to increase the number
of systems which can be studied at ultralow temperatures, including
atomic species \cite{Cold-atoms-review}, and also molecular
\cite{Carr-NJP-2009,Dulieu-Review-2011,Cote1997-mol,Cote1999-mol} and
ionic species \cite{Cote-review-ion,RMP-ion}.  In many cases, the
quantum regime where $s$-wave scattering dominates is still outside
the reach of today's experimental techniques, such as in atom-ion
hybrid system \cite{vladan09,kohl2010a,denschlag2010,kohl2010b,Ratschbacher-2013,Haze-2015,
Idziaszek-2011,Tomza-2015,Gacesa-Be-2017}. However, a large class of
physical systems is characterized by two states that are
asymptotically degenerate, and for which an initial scattering state
can be described by a superposition of those states; interference
between the two possible interaction paths may lead to resonant
exchange between the two states.  Such processes have been studied in
the scattering of neutral atoms, e.g., spin-flip in alkali atom
collisions \cite{Cote-Li-1994,Cote-Na-1994} with singlet and triplet potential
curves, as well as in $S$-$P$ excitation exchange for identical atoms
\cite{Bouledroua-2001}, and charge transfer between an atomic ion and
its neutral parent atom \cite{Cote-Dalgarno-2000,Cote-2000-mobility}.
Moreover, in cases involving quasi-resonant scattering, e.g., when
considering different isotopes, the resonant approximation adequately
describes the behavior of the system if the scattering energy is
higher than the energy splitting between the asymptotic states
\cite{Peng-Li,Peng-Be}.

In this Letter, we study the resonant exchange process
\begin{equation}
  X^\alpha + X^{\alpha'} \longrightarrow X^{\alpha'}+ X^\alpha,
\label{eq:exchange-process}
\end{equation}
where $\alpha$ and $\alpha'$ denote internal states.  For example, in
charge transfer ($X + X^+ \longrightarrow X^+ + X$) $\alpha$ denotes
the charge, with $\alpha=0$ and $\alpha'=+1$, while for excitation
exchange $\alpha=S$ and $\alpha'=P$ are the electronic states.  For
such resonant exchange processes, the cross section reads
\cite{Cote-Dalgarno-2000,Cote-review-ion,mott-massey}
\begin{equation}
   \sigma_{\rm exc} (E)= \frac{\pi}{k^2} \sum_{\ell =0}^\infty (2\ell +1) 
   \sin^2 (\eta_\ell^a - \eta_\ell^b ),
\label{eq:csec-def}
\end{equation}
where $k=\sqrt{2\mu E/\hbar^2}$ is the center of mass wave number for
the scattering of a pair of particle of reduced mass $\mu$ and
collision energy $E$. Here, $\eta_\ell^{a(b)}$ is the scattering phase
shift of the $\ell^{\rm th}$ partial wave along the interaction
potential $V_{a(b)}$, which correspond to the two asymptotically
degenerate channels

For energies high above the Wigner regime, 
where many partial waves are contributing, we can regard $\ell$ as a
continuous variable and  use the semi-classical expression
\cite{mott-massey,Harald-book}
\begin{equation} 
   \frac{\partial \eta_\ell}{\partial \ell} \approx \frac{\pi}{2} 
   + \int_{r_0(J)}^\infty dr \frac{\partial}{\partial J}
     \left[ 2\mu (E-V(r)) -\frac{J^2}{r^2} \right]^{1/2} \;,
\label{eq:eta-one}
\end{equation}
where $J=(\ell+\frac{1}{2})\hbar$, and $r_0(J)$ is the inner classical
turning point.  Although $\ell$ can be large, we assume that the
centrifugal term $J^2/r^2$ is a small perturbation on \color{black}
the potential $V(r)$, {\it i.e.}, $J^2/r^2 \ll 2\mu [E-V(r)]$. Under
such conditions, the scattering wave function still probes the inner
region, and $r_0$ depends weakly on $J$; we thus take it to be
independent of $J$ and equal to the $s$-wave turning point, i.e., $
r_0 (J) \equiv r_0$.  We now take the partial derivative out of the
integral (\ref{eq:eta-one}) and expand the integrand in small powers
of $J^2/r^2$ to obtain
\begin{equation}
 \frac{\partial \eta_\ell}{\partial \ell}  \approx  \frac{\pi}{2} 
 -J \int_{r_0}^\infty \frac{dr}{r^2} \frac{1}{\sqrt{2\mu (E-V)}}
 \equiv \frac{\pi}{2} -J \frac{A}{\hbar}\;, 
\label{eq:alpha-approx} 
\end{equation}
where $A$ is an integral independent of $J$\@.  Using
$J=(\ell+\frac{1}{2})\hbar$, we have $\partial \eta_\ell/\partial \ell
\approx \pi/2 + A (\ell + 1/2) + {\cal O} (\ell^2)$, which after
integration over $\ell$ gives
\begin{equation}
   \eta_\ell \approx \eta_0 + \frac{\pi}{2}\ell + \frac{A}{2} \ell (\ell +1) 
   + {\cal O} (\ell^3) \;.
\label{eq:eta-four}
\end{equation}
Using the Levinson theorem \cite{mott-massey,Harald-book}, we have $\eta_0 = N\pi +
\delta_0$, where $\delta_0$ is the $s$-wave phase shift modulo~$\pi$
and $N$ is the number of bound states.  Therefore, the phase shift
difference $\Delta\eta_\ell \equiv \eta_\ell^a - \eta_\ell^b$ reads
\begin{equation}
 \Delta \eta_\ell  \approx  \pi \Delta N\ + \Delta \delta_0 
                  + \ell (\ell +1) \frac{\Delta A}{2}   + {\cal O} (\ell^3) \;,
\label{eq:eta-diff}
\end{equation}
where $\Delta N = N_a-N_b$, $\Delta \delta_0 = \delta^a_0-\delta^b_0$, and 
$\Delta A = A_a - A_b$, and
 we obtain
\begin{equation}
   \sin^2(\Delta \eta_\ell) \approx \sin^2 \left[ \Delta \delta_0 
   + \ell (\ell +1) \frac{\Delta A}{2} \right]\;.
\label{eq:eta-diff-two}
\end{equation}

We now approximate the sum in Eq.~(\ref{eq:csec-def}) with an integral,
$\sigma_{\rm exc} \approx \frac{\pi}{k^2} \int_0^\infty d\ell (2\ell
+1) \sin^2 (\eta^a_\ell - \eta^b_\ell)$, which yields
   \color{black}
\begin{equation}
   \sigma_{\rm exc} \!\! \approx \!\!\frac{\pi}{k^2} \int_0^{L} \!\! d\ell (2\ell +1) 
   \sin^2 \!\!\left[\Delta \delta_0 \!+\!\ell (\ell +1) \frac{\Delta A}{2}\right]\!.
\end{equation}
In the integral above, the upper limit $L$ is set to a suffieciently
large value of $\ell$ such that the two phase shifts become equal;  thus,
there will be no further contribution to the integral for
$\ell>L$.
This occurs when the centrifugal barrier becomes dominant
for both potential curves \cite{Cote-Dalgarno-2000}.  Changing
variable to $x\equiv \Delta \delta_0 +\ell (\ell +1) \frac{\Delta
  A}{2}$, our integral simply becomes $\sigma_{\rm exc} \simeq
\frac{\pi}{k^2} \frac{2}{\Delta A}\int_{x_0}^{x_L} dx\;\sin^2 x =
\frac{\pi}{k^2} \frac{2}{\Delta A} \left[ \frac{x}{2} -
  \frac{1}{4}\sin (2x) \right]_{x_0}^{x_L}$, with $x_0=\Delta
\delta_0$ and $x_L = \Delta \delta_0 + L(L+1)\Delta A/2$, giving
\begin{eqnarray}
   \sigma_{\rm exc} & \simeq &  \frac{\pi}{k^2} \frac{1}{\Delta A} \left[ L(L+1)\frac{\Delta A}{2} 
      + \frac{1}{2}\sin (2 \Delta \delta_0)\right. \nonumber \\
  & & \left. \hspace{.45in} - \frac{1}{2}\sin (2 \Delta \delta_0 + L(L+1)\Delta A) 
          \right].
\label{eq:csec-app-1}
\end{eqnarray}

Finally, we can simplify our result if we employ the approximation
$L(L+1)\Delta A \ll 1$, which can be justified if we examine the
parameter $A_i$ given by Eq.(\ref{eq:alpha-approx}), {\it i.e.}
\begin{equation}
   A_i = \frac{\hbar}{\sqrt{2\mu}} \int_{r^i_0(E)}^\infty 
             \frac{dr}{r^2} \frac{1}{\sqrt{E-V_i(r)}} \; ,
\label{eq:A-def}
\end{equation}
where the inner turning point $r^i_0$ for the potential $V_i$ depends
on the scattering energy $E$. In resonant processes where the
long-range tail of each $V_i$ is the same, only their shorter range
difference contribute to $\Delta A$. Typically, $\Delta A$ varies
little with $E$, and is of the order 0.01--0.001 for the physical
systems considered in this Letter, with $\Delta A$ smaller for heavier
systems due the $2\mu$ factor.  We now return to
Eq.~(\ref{eq:csec-app-1}) and we use the approximation $\sin [2 \Delta
  \delta_0 + L(L+1)\Delta A ]\approx L (L+1)\Delta A \cos (2 \Delta
\delta_0)+\sin (2 \Delta \delta_0)$ to obtain
\begin{eqnarray}
   \sigma_{\rm exc}(E) & \simeq & 
   \frac{\pi}{k^2} \frac{1}{\Delta A}  L (L+1) \frac{\Delta A}{2} 
        [ 1- \cos (2 \Delta \delta_0)] \;, \nonumber \\
  & \approx & \frac{\pi}{k^2} L^2 
        \sin^2 \Delta \delta_0 (E)\;,
\label{eq:csec-app-2}
\end{eqnarray}
where we assume $L(L+1)\approx L^2$.

The expression above can be related to the Langevin cross section
$\sigma_L$.  Indeed, we defined $L$ as the maximum $\ell$ for which
the phase shift difference is non-negligible, which corresponds to the
height of the centrifugal barrier slightly larger than $E$\@.  This
critical value of $\ell$ also defines $\sigma_L$, which is determined
by the impact parameter $b_{\rm max}$ still allowing penetration in
the inner region where the exchange process occurs with unit
probability \cite{mott-massey,Cote-Dalgarno-2000,Cote-review-ion}.   With the
impact parameter $b\equiv (\ell + \frac{1}{2})/k$, we obtain
\begin{equation}
   \sigma_L (E)= \pi b_{\rm max}^2 \simeq \frac{\pi}{k^2} L^2 \;, 
\label{eq:csec-Langevin-def}
\end{equation}
where we assume $L+\frac{1}{2} \approx L$ for large $L$.  We remark
that $L$ has the same value for both potentials $V_a$ and $V_b$, which
is a valid assumption for the energy range dominated by the long range
tail (which is the same for both potentials).  For potentials with an
asymptotic behavior $V(r)\sim -C_n/r^n$, the location of the top of
the barrier is $r_{\rm top} = \left(\frac{\mu n C_n}{\ell
  (\ell+1)\hbar^2}\right)^{\frac{1}{n-2}} $, and $L(E)$ is obtained
from $E=V (r_{\rm top})$, which yields
\begin{equation}
   L(L+1) =  \frac{1}{\hbar^2} \left( \frac{n}{n-2}\right)^{\frac{n-2}{n}}
          (\mu n C_n)^{\frac{2}{n}} (2\mu E)^{\frac{n-2}{n}}\;.
   \label{eq:L(L+1)}       
\end{equation}
Again, assuming $L(L+1)\approx L^2$, we can write the Langevin cross section 
$\sigma_L  = \frac{\pi}{k^2} L^2$ as
\begin{equation}
   \sigma_L (E)=  \pi
                  \left( \frac{n}{n-2}\right)^{\frac{n-2}{n}}
                  (n C_n)^{\frac{2}{n}} (2 E)^{-\frac{2}{n}} \;.
   \label{eq:Langevin-general}
\end{equation}
The expressions for the most common long range inverse power-law potentials
are listed in Table~\ref{tab1}, with $n=3$ appearing in dipole allowed
excitation exchange, $n=4$ in polarization potentials between atoms
and ions, and $n=6$ in van der Waals interactions between ground state
atoms.

\begin{table}
\caption{Langevin cross section $\sigma_L$ for various $n$.}
\label{tab1}
\begin{tabular}{r|cccccc} \hline\hline %&&&&&&\\
 $n$ && 3 & & 4 & & 6  \\ \hline
  %&&&&&\\ 
$ \sigma_L $ &  &$\displaystyle 3 \pi  \left(\frac{ C_3}{2E}\right)^{2/3}$ 
   && $\displaystyle 2\pi \left(\frac{C_4}{E}\right)^{1/2}$
   && $\displaystyle \frac{3\pi}{2}\left(\frac{2 C_6}{E}\right)^{1/3}$ \\ 
  %&&&&&&\\ 
  \hline\hline
\end{tabular}
\end{table}

The simple expression $\sigma_{\rm exc} = \sigma_L \sin^2 (\Delta
\delta_0)$ is obtained by combining Eqs.(\ref{eq:csec-Langevin-def})
and (\ref{eq:csec-app-2}).  However, for energies low enough that
$L^2<1$, $\sigma_{\rm exc}$  will decrease rapidly and not capture the
asymptotically constant $s$-wave cross section of the Wigner regime as
$E\rightarrow 0$. This is remedied by adding the low energy
contribution, so that we finally can write the resonant exchange cross
section as
\begin{equation}
  \sigma_{\rm exc} (E)= \left[\frac{\pi}{k^2}+\sigma_L(E) \right] \sin^2 \Delta \delta_0 (E).
  \label{eq:csec-final}
\end{equation}
This equation explicitly shows how the $s$-wave regime modulates the
Langevin cross section, leading to a signature of the $s$-wave regime
at higher temperatures.

To illustrate the effect of the $s$-wave regime at higher energies, we
first consider charge transfer between Yb and Yb$^+$ for a variety of
isotopes. As noted in a previous article on resonant charge transfer
\cite{Peng-Yb-2009}, the cross section exhibits an intermediate ``modified"
Langevin regime where the $\sigma_{\rm exc}$ seems to be affected by
the ultracold behavior, even if many partial waves contribute to its
overall value. In fact, one could notice a ``correlation" between the
cross section at ultralow energy and at much higher energies. The
expression above provides the explanation for the correlation noted in
\cite{Peng-Yb-2009}: if the $s$-wave phase shifts corresponding to states
$a$ and $b$ happen to have nearly equal values, their
  difference remains small even at higher scattering energy.  This
can be seen from the WKB approximation, or equivalently
Eq.(\ref{eq:eta-four}), as $\eta_\ell$ varies slowly with $\ell$.
  If $\Delta\eta_0$ is small, the phase difference for
higher $\ell$ will also remain small for a wide
range of partial waves, due to the phase shift
``locking" described above, and will result in a reduced cross
section.  Naturally, the applicability of Eq.(\ref{eq:csec-final})
depends on the details of the potentials and validity of the
approximations involved in its derivation.

\begin{figure}
\includegraphics[width=1\linewidth]{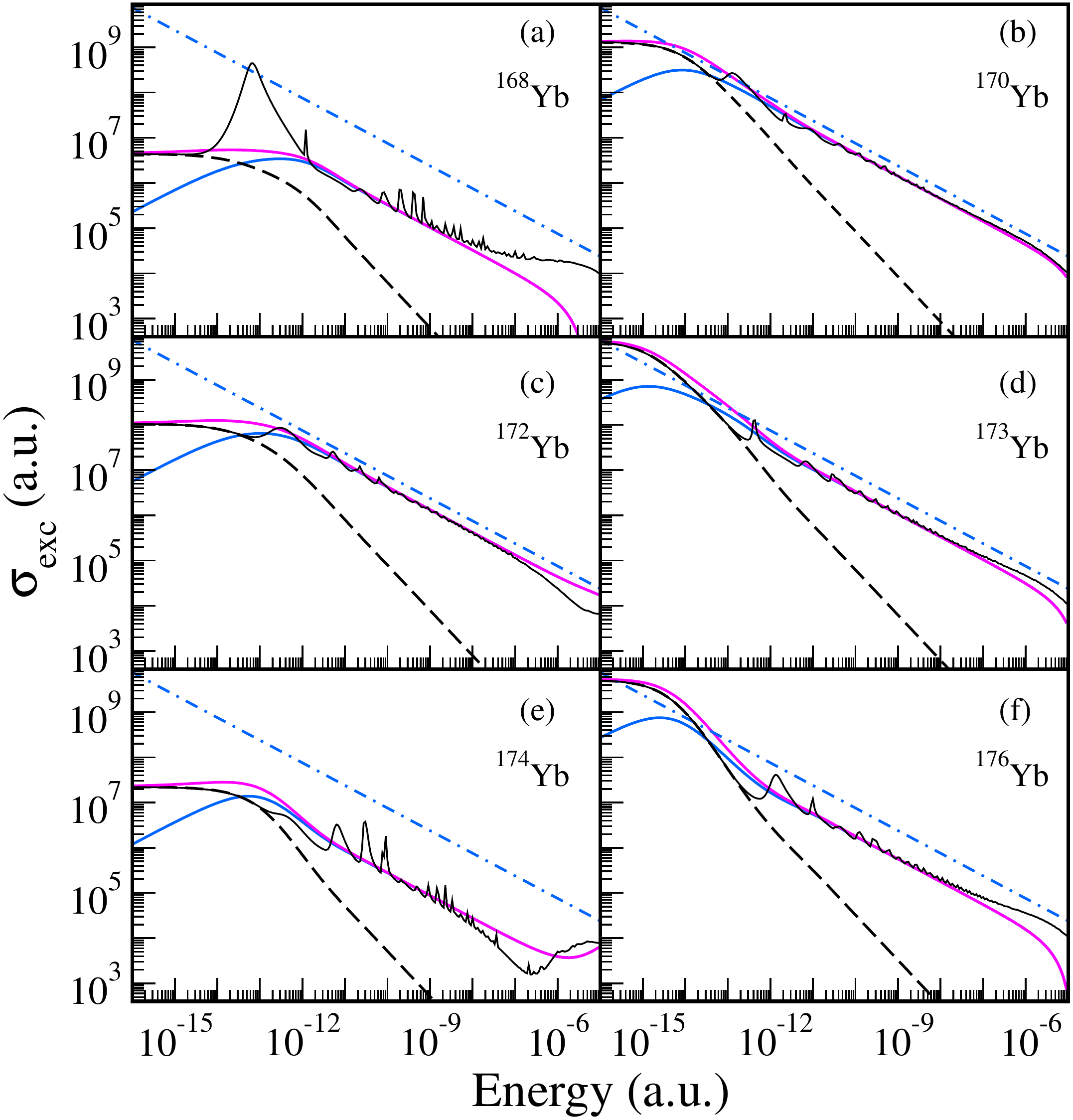}
\caption{(Color on line) Resonant charge transfer $\sigma_{\rm exc}$
  between various isotopes of Yb and Yb$^+$ vs. scattering energy
  $E$. The numerical results (black line) are compared to
  Eq.(\ref{eq:csec-final}) (magenta line), together with its components;
  the $s$-wave contribution $\frac{\pi}{k^2}\sin^2\Delta\delta_0$
  alone (black dashed line) and the $\sigma_L \sin^2\Delta\delta_0$
  alone (solid blue line). The standard $\sigma_L$ (blue dot-dashed
  line) is shown for comparison purposes. Isotopes 168 in (a), 174 in
  (e), and 176 in (f) show significant suppression when compared to
  $\sigma_L$, while 170 in (b), 172 in (c), and 173 in (d),
  $\sigma_{\rm exc}\approx \frac{1}{2}\sigma_L$ over a wide range of
  energies.  }
\label{fig:Yb}
\end{figure}

In Fig.~\ref{fig:Yb}, we compare the simple expression 
(\ref{eq:csec-final}) to the full numerical results computed using the
 approach  described in \cite{Peng-Yb-2009}. The
potentials $V_g$ and $V_u$ corresponding to the $^2\Sigma_g^+$ and
$^2\Sigma_u^+$ of Yb$_2^+$ behave as $-C_4/r^{-4}$ with $C_4= 72.5$
a.u. at large separation.  The cross section $\sigma_{\rm
  exc}(E)$ depends strongly on the atomic mass of the Yb isotopes.
 In each plot of Fig.~\ref{fig:Yb}, the ``standard"
Langevin cross section $\sigma_L$ is included to
emphasize the effect of the $s$-wave phase shifts.
In some cases, like plots (b), (c), and (d), corresponding to isotopes
170, 172, and 173, both $\sigma_L$ and $\sigma_{\rm exc}$ give similar
values, {\it i.e.} the $s$-wave has no sizable effect on the cross
section. Actually, $\sigma_{\rm exc}$ is roughly
$\frac{1}{2}\sigma_L$, which is to be expected in general since
$\langle \sin^2\Delta\delta_0\rangle = 1/2$ if the phase shift
difference $\Delta \delta_0$ is random.  However, in other cases, like
for isotopes 168, 174, or 176 in (a), (e) and (f) respectively, the
signature of the s-wave regime is noticeable, with a
reduction of two orders of magnitude for (a) and (e), and one for
(f). As mentioned above, this is due to the accidental proximity in
values of the residual phase shifts $\delta_0^{g(u)}$ corresponding to
$V_{g(u)}$, and the phase shift locking as $E$ increases. We note that
according to Eq.(\ref{eq:L(L+1)}), $L^2<1$ when $E$ becomes smaller
than roughly $10^{-13}$ a.u.  for the Yb systems above, at which point
the $s$-wave contribution (negligible at higher $E$) satisfying the
Wigner regime kicks in. It is worth noting that when the $s$-wave
suppression of $\sigma_{\rm exc}$ is significant, as in Figs.~\ref{fig:Yb} (a)
and (e), the underlying shape resonances become more apparent as the
background cross section diminishes. Naturally, these resonances are
absent from our WBK-treatment in Eq.(\ref{eq:csec-final}), which
reproduces the general trend of the
numerical results  over a large range of $E$.

\begin{figure}
\includegraphics[width=1\linewidth]{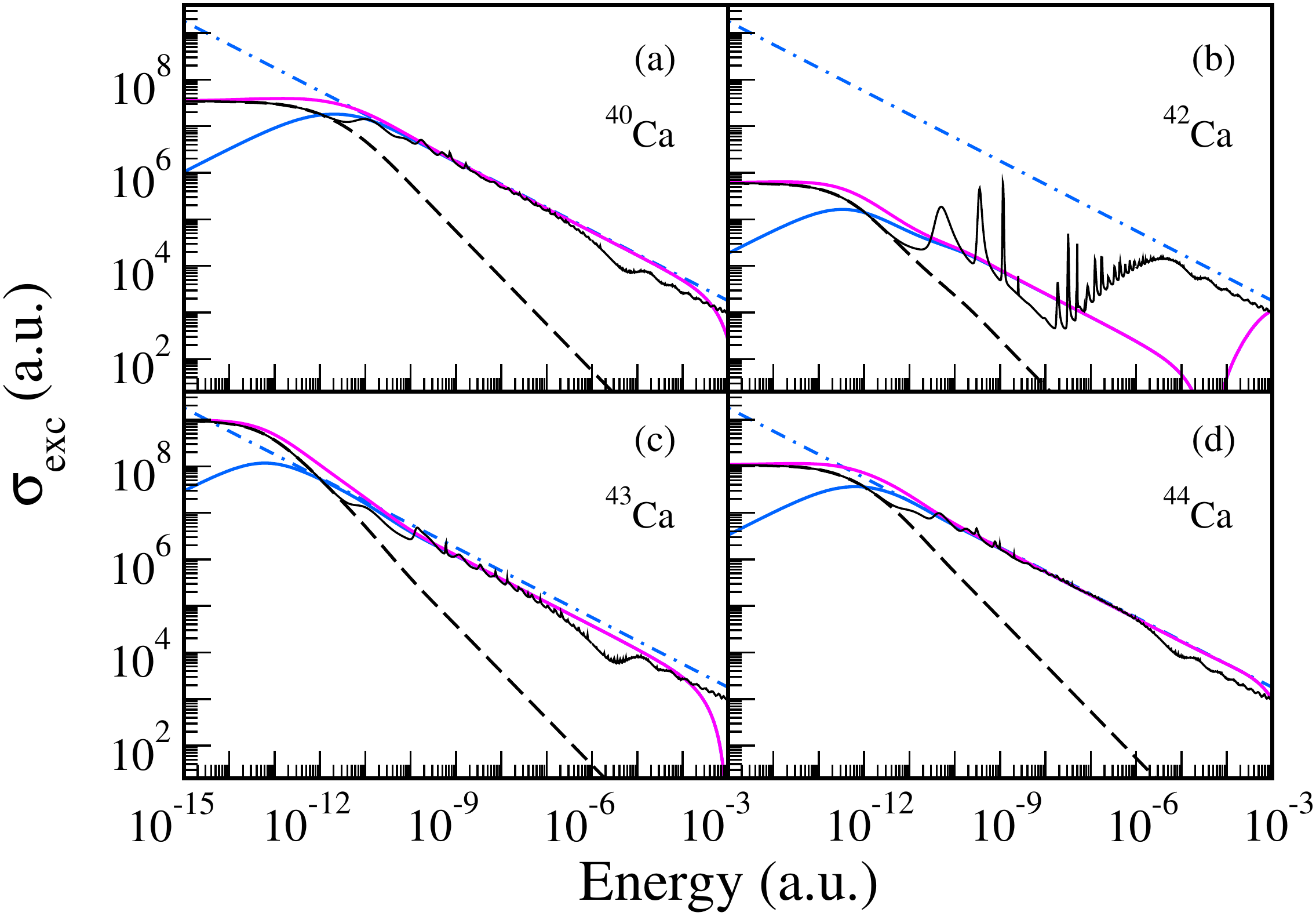}
\caption{(Color on line). Same as Fig.~\ref{fig:Yb} for spin-flip in
  collision of Na and various isotopes of Ca$^+$, 40 in (a), 42 in
  (b), 43 in (c), and 44 in (d). Significant suppression occurs in
  (b), while $\sigma_{\rm exc}$ is close to $\sigma_L$ for the other
  isotopes.  }
\label{fig:NaCa}
\end{figure}

Atom-ion scattering can also lead to a resonant spin-flip process,
such as in \cite{Makarov-2003,Smith2014}, where a ground state Na atom approaching
a Ca$^+$ ion can interact via a singlet A$^1\Sigma^+$ or a triplet
a$^3\Sigma^+$ electronic state described by the singlet (triplet)
potential $V_{S(T)}$ with corresponding residual phase shift
$\delta_\ell^{S(T)}$. In that work, $\sigma_{\rm exc}$ was found to be
roughly $\frac{1}{2}\sigma_L$.  Recent experiments on Yb$^+$+$^{87}$Rb \cite{Ratschbacher-2013},   
Yb$^+$+$^6$Li \cite{Tomza-2018}, and $^{88}$Sr$^+$+Rb \cite{Sikorsky-2018,Timur-arXiv-2018} 
have explored spin-flip dynamics.
In Fig.~\ref{fig:NaCa}, we explore the
effect of the $s$-wave scattering on the spin-flip in Ca$^+$+Na, using
$V_{S(T)}$ described in \cite{Makarov-2003,Gacesa-NaCa-2016} (behaving as $r^{-4}$ at
large $r$) for four isotopes of Ca, namely 40, 42, 43, and 44. Again,
the simple expression (\ref{eq:csec-final}) agrees with the numerical
cross sections over a wide range of energy. Fig.~\ref{fig:NaCa} shows
a variety of behavior of $\sigma_{\rm exc}$. For example, in (a) and
(d), $\sigma_{\rm exc}\approx \sigma_L$ at higher energies,
corresponding to $\Delta\delta_0 =\delta_0^S-\delta_0^T \approx
\pi/2$, while (c) depicts a small suppression by a factor of about
1.5. The case of $^{40}$Ca leads to a substantial reduction of about
200, again revealing the underlying shape resonances.

\begin{figure}
\includegraphics[width=1\linewidth]{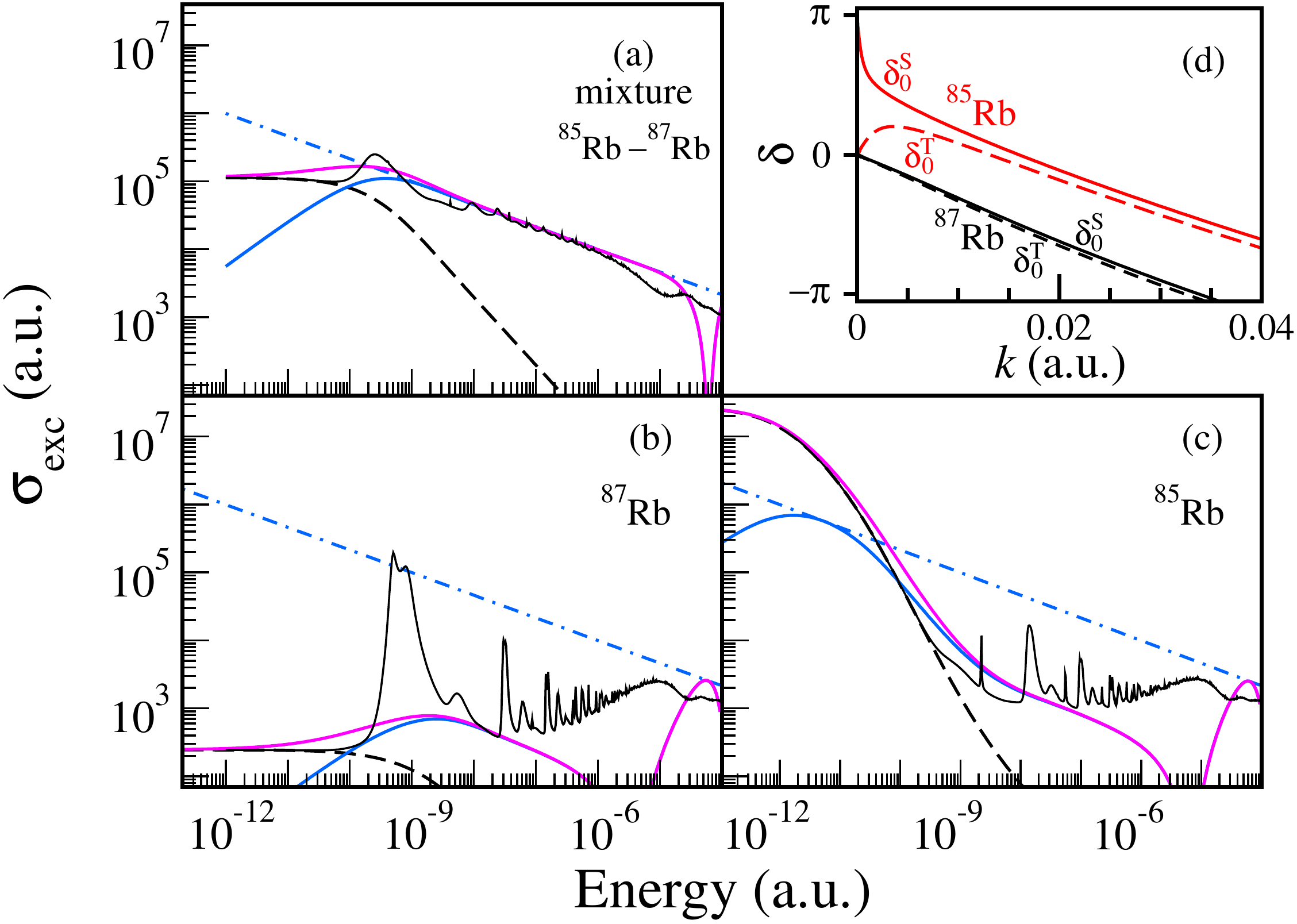}
\caption{(Color on line). Same as Fig.~\ref{fig:Yb} for spin-flip in
  collision Rb atoms, with $^{85}$Rb+$^{87}$Rb mixture in (a), pure
  $^{87}$Rb in (b), and pure $^{85}$Rb in (c). Both pure cases show
  extreme suppression compared to $\sigma_L$, with resonances revealed
  by the small background.  Corresponding $\delta_0^{S(T)}$ are
  depicted in (d). }
\label{fig:Rb}
\end{figure}

Spin-flip collisions have also been studied between neutral atoms,
especially alkali atoms, such as in Li \cite{Cote-Li-1994} or Na
\cite{Cote-Na-1994}. Again, the ground state atoms approach each other in
a superposition of singlet X$^1\Sigma_g^+$ and triplet a$^3\Sigma_u^+$
states, behaving asymptotically as $-C_6/r^6$.  To illustrate the
effect of the $s$-wave regime on $\sigma_{\rm exc}$, we consider a
system for which the scattering lengths are known to be close to each
other, namely $^{87}$Rb. Using the potential curves described in
\cite{Timur-Rb2-2008}, we computed $\sigma_{\rm exc}$ for pure
$^{87}$Rb, $^{85}$Rb, and their mixture. The results are shown in
Fig.~\ref{fig:Rb}; $\sigma_{\rm exc}$ for the mixture in (a) follows
roughly $\sigma_L$ away from ultracold temperatures. As expected, for
$^{87}$Rb in (b) with both singlet and triplet scattering lengths
almost equal %($a_S=98.9$ a.u. and $a_T= 90.1$ a.u.)
($a_S\approx a_T\approx 100$~a.u.), the $s$-wave
suppression is drastic, with shape resonances emerging from the
suppressed background.  Although not perfect, the simple expression
(\ref{eq:csec-final}) tracks the overall reduction of a factor of
$10^4$ in $\sigma_{\rm exc}$.  Much more surprising is the result for
$^{85}$Rb (c) with scattering lengths
%($a_S=2537.8 $ a.u. and $a_T=-389.5$ a.u.)
($a_S\approx 2500$ a.u.\ and $a_T\approx -390$~a.u.)
which are very different.  In this case, one could
have expected the system to follow the Langevin case. However, a
closer look at the $s$-wave phase shifts explains the seemingly
unusual cross section.  The large positive and negative values of
$a_S$ and $a_T$ imply rapid changes of $\delta_0^{S(T)}$ with energy
in the Wigner regime, as shown in Fig.~\ref{fig:Rb}(d). The large
initial value of the phase shift difference $\Delta\delta_0$ (mod
$\pi$) quickly evolves into a much smaller value, comparable to the
case of $^{87}$Rb.

As a final example, we consider a system for which the interaction
potentials behave as $r^{-3}$ at long range. Many examples occur in
nature, such as excitation exchange \cite{Bouledroua-2001}, or in the
scattering of metastable atoms, like H(2s)+H(2s)
\cite{Forrey-PRL-2000,Jonsell-PRA-2002}, or the excitation exchange in
metastable helium He(1$^1$S)+He$^*$(2$^3$P)
\cite{Vrinceanu-2010,Peach-2017}.  Here, we focus our attention on
Cs$^+$+Cs$(6p)$ which can lead to the exchange of the $6p$ excitation
onto Cs$^+$. Four excited electronic states are involved if we neglect
spin-orbit coupling, two $\Sigma^+_{g(u)}$ and two $\Pi_{g(u)}$, each
correlated to the Cs$_2^+(6p)$ asymptote, and described by potential
curves $V^\Sigma_{g(u)}$ and $V^\Pi_{g(u)}$ and corresponding residual
phase shifts $\delta_{\Sigma,\ell}^{g(u)}$ and
$\delta_{\Pi,\ell}^{g(u)}$. The cross section is \cite{mott-massey,Bouledroua-2001}
\begin{equation}
   \sigma_{\rm exc} = \frac{\pi}{3k^2} \sum_{\ell =0}^\infty (2\ell +1)
   \left[\sin^2 \Delta \delta_\ell^\Sigma 
                             +   2\sin^2\Delta\delta_\ell^\Pi \right],
\end{equation}
where $\Delta \delta_\ell^\Sigma \equiv \delta_{\Sigma,\ell}^{g} -
\delta_{\Sigma,\ell}^{u}$ and $\Delta \delta_\ell^\Pi \equiv
\delta_{\Pi,\ell}^{g} - \delta_{\Pi,\ell}^{u}$.  Since the $\Sigma$
and $\Pi$ curves have different $C_3$ values, $L$ for both sets is
different. Using our approximations, $\sigma_{\rm exc}$ becomes
\begin{equation}
   \sigma_{\rm exc} \!=\! \frac{1}{3}\! \left[\frac{\pi}{k^2} \!+\!  \sigma_L^\Sigma \right]\! \sin^2\! \Delta \delta^\Sigma_0
    \!+\! \frac{2}{3}\! \left[\frac{\pi}{k^2} \!+\! \sigma_L^\Pi \right]\! \sin^2 \! \Delta \delta^\Pi_0,
\end{equation}
where $\sigma_L^{\Sigma(\Pi)}$ is obtained with the appropriate value of $C_3$.

\begin{figure}[b]
\includegraphics[width=1\linewidth]{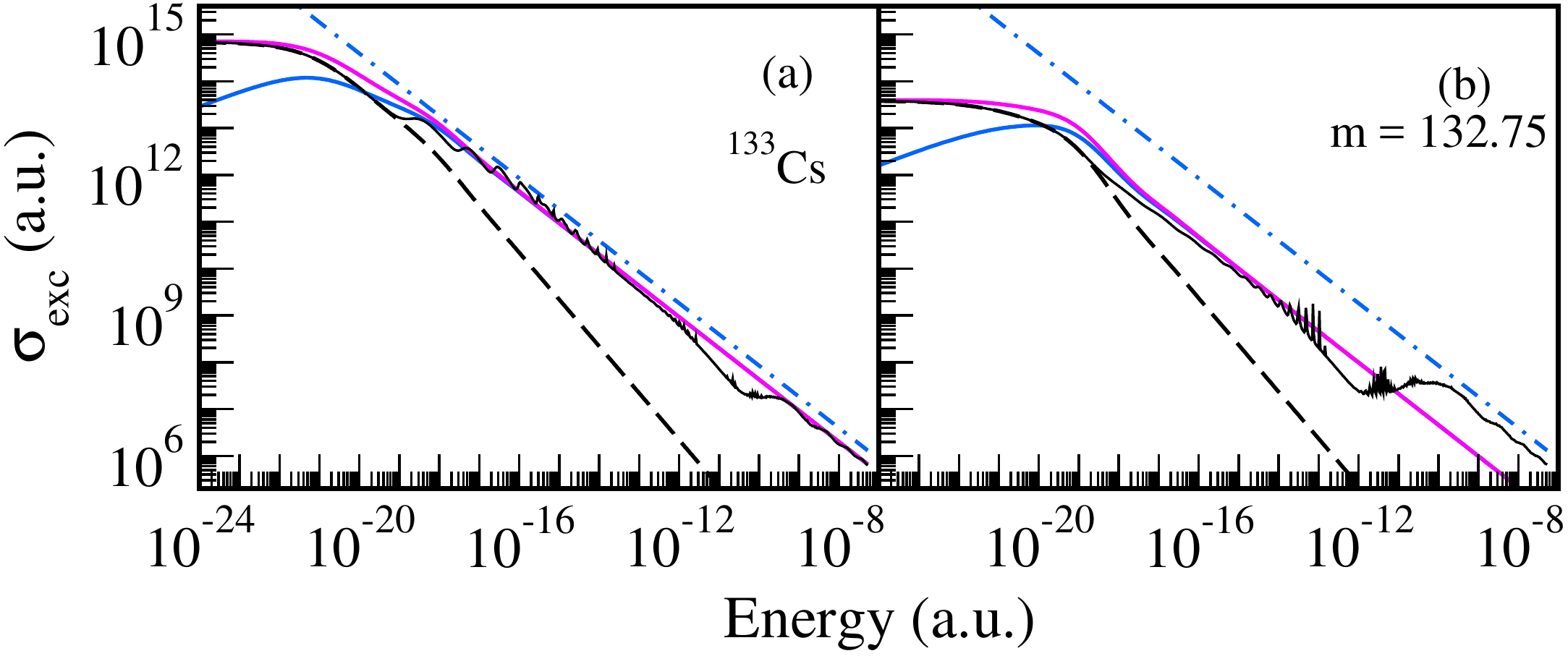}
\caption{(Color on line). Same as Fig.~\ref{fig:Yb} for excitation
  exchange in Cs$^+$+Cs($6p$), for the real mass in (a), and a
  fictitious mass of 132.75 u in (b), to illustrate the $s$-wave
  suppression.  }
\label{fig:Cs}
\end{figure}

The results shown in Fig.~\ref{fig:Cs} were obtained with the
$^2\Sigma^+_{g(u)}$ and $^2\Pi^+_{g(u)}$ from Jraij {\it et al.}
\cite{Jraij-Cs2}. The $\Pi$ curves are repulsive at large
separation behaving as $+C_3^\Pi/r^3$ with $C_3^\Pi = -13.95$ a.u.;
the {\it gerade} and {\it ungerade} phase shifts are basically equal
for all $\ell$, their cancellation leading to a negligible $\Pi$
contribution. The two $\Sigma$ curves are attractive, and were matched
at large separation to $-C_4/r^4 -C_3/r^3$ with $C_4=1082$ a.u., and
$C_3^\Sigma = 27.9$ a.u. Since there is only one stable isotope of
Cs, we rescaled its mass to model a different isotope. For the real mass
of Cs, the cross section is roughly half the Langevin cross section,
while choosing $m_{\rm Cs} = 132.75$~u, the cross section is reduced
by a factor of 20, again exposing the resonances as in previous
examples.

In conclusion, we derived a simple expression for resonant scattering
processes, relating the cross section to the Langevin cross section
and the $s$-wave regime.  By relying on the WKB approximation, we
derived the expression for the exchange cross section, and showed that
it has wide applicability, as long as the pairs of potential curves
have the same long range tail.  We illustrated the range of
applicability using various resonant systems such as charge transfer,
spin-flip, and excitation exchange, and for a variety of long range
inverse power-law tail behaving as $r^{-n}$ covering the most common
powers. The expression points to the signature of the $s$-wave regime
at higher temperatures, and how the $s$-wave phase shift ``locking"
actually modulates the cross section. The results presented here also
provide a diagnostic tool particularly relevant to system for which
ultracold temperatures are not easily achievable, such as atom-ion
hybrid systems for which the nK regime remains a challenge.  In fact,
by measuring the cross section or rate for a resonant process, 
  e.g.,  charge transfer or spin-flip, at higher temperatures more
easily accessible, one can gain information about the $s$-wave
regime. If a sizable suppression is observed as compared to
$\sigma_L$, this implies that the $s$-wave phase shifts are close to
each other. In addition, the suppression helps revealing shape
resonances otherwise submerged which, together with the $s$-wave
suppression, can help determining the potential curves more accurately
down to the $s$-waves.  Finally, the expression should be applicable
to quasi-resonant processes as well \cite{mott-massey}, such as charge
transfer in systems with mixed isotopes \cite{Peng-Li, Peng-Be}, or in
reactions involving isotope substitutions, as long as the scattering
energy is larger than the energy gap between the asymptotes of the
relevant potentials.

\begin{acknowledgments}
This work was partially supported by the National Science Foundation
Grant PHY-1415560 (IS)
and by the MURI US Army Research Office Grant No. W911NF-14-1-0378 (RC).
\end{acknowledgments}

\bibliography{langevin-ref}

\end{document}